# Stream Clustering using Probabilistic Data Structures

Andrei Sorin Sabau,
Faculty of Mathematics and Computer Science, University of Pitesti, Str. Targul din Vale, nr.1, Pitesti 110040, Romania. Contact: andrei@asabau.com.

**Abstract**. Most density based stream clustering algorithms separate the clustering process into an online and offline component. Exact summarized statistics are being employed for defining micro-clusters or grid cells during the online stage followed by macro-clustering during the offline stage. This paper proposes a novel alternative to the traditional two phase stream clustering scheme, introducing sketch-based data structures for assessing both stream density and cluster membership with probabilistic accuracy guarantees. A count-min sketch using a damped window model estimates stream density. Bloom filters employing a variation of active-active buffering estimate cluster membership. Instances of both types of sketches share the same set of hash functions. The resulting stream clustering algorithm is capable of detecting arbitrarily shaped clusters while correctly handling outliers and making no assumption on the total number of clusters. Experimental results over a number of real and synthetic datasets illustrate the proposed algorithm quality and efficiency.

## 1. Introduction

A typical technology ecosystem is now producing more data than the available storage space allocated to it. Additional space and time restrictions apply, rendering traditional data mining techniques, including clustering, inefficient. Compared to traditional clustering, data stream clustering can only perform a single pass over the data while maintaining summaries and adapting to the underlying concept drift and shift. Most real world data sets and data streams exhibit arbitrarily shaped clusters accompanied by various degree of noise or outliers. Among the plethora of existing stream clustering algorithms, density micro-clustering and density grid based algorithms meet, with a large percentage of success, the above criteria. Most such algorithms follow the two phase stream clustering scheme [1] employing an online and offline component.

BloomStream, the proposed density based clustering algorithm, relies on sketch-based data structures for assessing both stream density and cluster membership. A count-min sketch [2] using a damped window model estimates stream density. Bloom filters [3] employing a variation of active-active buffering [4] estimate cluster membership. Instances of both types of sketches share the same set of hash functions generated from linear combinations of just two independent, uniform random hash functions without any loss in the asymptotic failure probability.

The rest of the paper is organized as follows. In Section 2, current state of knowledge is reviewed for grid and sketch based stream clustering algorithms. Section 3 introduces the main concepts underlying the proposed algorithm. Section 4 contains its theoretical analysis and algorithmic details. Section 5 reports the performance study conducted on both real and synthetic data sets. Section 6 concludes the paper.

## 2. Related work

This section reviews related work on data stream clustering focusing on grid and sketch based summarization techniques as the main building blocks of the proposed algorithm. Both techniques achieve low space and computation complexity by approximating the data stream. The former maintains generalized information in terms of discretized intervals while the latter records frequency counts with probabilistic accuracy guarantees.

### 2.1. Grid based summarization techniques

Grid based techniques map an unlimited length data stream into a finite number of grid cells by partitioning the data space into ranges along each dimension. The number of stream instances within a cell determines its density. Dense cells are aggregated into dense regions forming clusters. This density based approach allows for arbitrarily shaped cluster identification and outlier detection over a single data scan.

D-Stream [5] maintains a characteristic vector for each grid cell containing, among others, cell density and last update time. Based on user defined density thresholds coupled with a density decay factor each cell is labeled as dense, transitional or sporadic. Dense and transitional cells are subject to density-based clustering while sporadic ones indicate outliers. Periodic pruning is conducted in order to remove the latter. DD-Stream [6] extends D-Stream by storing additional centroid-like information to the characteristic vector. This allows for extraction and assignment of boundary points to the nearest neighboring grid cell center with improved clustering results. D-Stream II [7] adds a grid attraction vector as a non-symmetric measure to D-Stream characteristic vector reflecting to what extent data in a given grid cell is closer to its neighbors. Two grid cells are said to be strongly correlated if their attraction in both directions exceed a user defined threshold. The clustering procedure departs from standard density based clustering as only strongly correlated grid cells are being merged.

The above algorithms employ a user-defined, fixed grid structure while following the two phase stream clustering scheme. This leads to predictable and constant per-point processing time, a highly desirable characteristic in stream clustering, but leaves no instruments for addressing data locality and dimensionality challenges inherited from traditional grid clustering algorithms.

Cell-Trees [8] adapts to data locality by using a dynamic discretization process. It uses cell tree, a multi-dimensional search tree similar to a kd-tree, to recursively partition cells based on their local density. The cell tree maximum depth equals the stream data dimensionality with a different dimension being used for splitting at each level. A multi-dimensional grid cell is thus represented by a cell tree path allowing for a hierarchical clustering of the stream data. MR-Stream [9] also uses a hierarchical structure but here a cell is divided not across a single dimension but across all dimensions resulting in up to $2^d$ direct sub-cells with $d$ being the stream data dimensionality. Clustering at multiple resolutions can be conducted in a manner similar to D-Stream.

PKS-Stream [10] adapts to high dimensional data streams by storing the relative position between non-empty grid cells. This removes the high computational cost of neighbor search when clustering. An unbalanced, fixed height Pks-tree is introduced containing grid cells of increased granularity depending on tree level. Only K-cover grid cells are stored represented by cells at either maximum granularity or having more than K-1 K-cover cells as siblings. Clustering is being conducted on leaf nodes with K-covers revealing non-empty neighboring grids.

*2.2. Sketch based summarization techniques*

Sketch based techniques employ non-cryptographic hash functions for approximating frequency counts over the data stream. Since all frequency counts are non-negative, the error due to hash cell collisions is one sided with a probabilistic upper bound guarantee of the overestimation. With constant query and update time, sketches are providing the data the clustering process will be conducted on.

Partition based CSketch [11] approximates dot product similarity between data stream instances and cluster centroids. Based on the same set of hash functions, the algorithm maintains a separate count-min sketch for each cluster. Each sketch table contains frequency counts of all attribute values for data instances assigned to the corresponding cluster. When clustering a new data instance, corresponding frequency counts are being retrieved and used to approximate the dot product across all existing clusters. The cluster yielding the largest dot product is selected for assignment. It is important to note the algorithm can only handle categorical data since count-min computes the frequency count of all distinct values encountered.

XStreamCluster [12] conducts two steps clustering of streaming XML documents. The latter uses a revised Jaccard similarity measure based on XML graph edges. Documents are represented as directed graphs with all edges being hashed and approximated into bloom filters. Clusters are constructed by merging bloom filters of the assigned documents using bitwise-OR operations. Document edge count, cluster edge count and document - cluster edge intersection used for computing Jaccard similarity can all be estimated from the maintained bloom filters.

UESStream [13] uses a modified count-min sketch to summarize uncertain data streams. It uses sketch*-metric [14] to estimate the similarity between any two data streams by only using stream sketches. The selected similarity measure is the Kullback-Leibler divergence, used for measuring the statistical difference between the data streams. Partitioned based clustering is conducted using a k-means variant.

**3. Sketch based model**

BloomStream applies sketches for both density and cluster membership estimation. With grid coordinates as input the former uses count-min point queries, the latter bloom filter signatures. To speed up computation both probabilistic data structures use the same set of hash functions over equal length hash tables. This level of compatibility can only be obtained by modifying the original sketches with the implications thoroughly discussed from a set / multiset perspective.

*3.1. Bloom filter*

A standard Bloom filter supports approximate set membership queries. Although the space savings come at the cost of false positives, an advantageous trade-off can often be obtain by keeping the probability of error at a sufficiently low rate.

Let $U$ be the universe of possible set elements and $S \subset U$ a set with $n = |S|$ elements. $S$ is represented by using a bit vector $B$ of $m = O(n)$ bits, initially all set to 0. Mapping elements from $U$ to $B$ is achieved via $k$ independent uniformly random hash functions $h_1, \dots, h_k: U \to \{1, \dots, m\}$.

For each element $x \in S$, $k$ bits from $B$ are set to 1 at positions $h_1(x), \dots, h_k(x)$. Given an element $q \in U$, if $\forall 1 \leq i \leq k, B(h_i(x)) = 1$, the element is likely to be contained in the set, otherwise it is certainly not contained. The above condition can hold true even for $q \notin S$ if elements from $S$ are mapped into the same bit positions as $q$. This false positive rate or bloom error is dependent on the selection of parameters $m$, $k$ and the cardinality of set $S$. The probability that a bit is still 0 after hashing the entire set $S$ into the bloom filter is:

$$\left(1 - \frac{1}{m}\right)^{kn} \approx e^{-\frac{kn}{m}} \quad (1)$$

The false positive probability *fp* of having all $k$ bit positions set to 1, erroneously claiming $q \in S$, becomes:

$$fp = \left(1 - \left(1 - \frac{1}{m}\right)^{kn}\right)^k \approx \left(1 - e^{-\frac{kn}{m}}\right)^k \quad (2)$$

Designing $k$ different independent hash functions and computing the corresponding hashes for each element to be inserted is prohibitively expensive in a data stream scenario. Fortunately the requirement for independent hash functions can be relaxed. Linear combinations of just two independent, uniform random hash functions $h_1(x)$ and $h_2(x)$ are used to generate the remaining ones while maintaining the same asymptotic false positive probability [15]. Given a prime $p$, $1 \leq i \leq k$, we define the hash functions $g_i(x)$ as:

$$g_i(x) = h_1(x) + i h_2(x) \bmod p \quad (3)$$

The above holds if we separate the range of each hash function. With a bit vector size of $m = kp$ bits, each hash function $h_1, \dots, h_k: U \to \{1, \dots, p\}$ gets allocated $m/k$ consecutive bit locations as disjoint ranges. Asymptotically the probability that a bit is still 0 after hashing the entire set $S$ remains the same:

$$\left(1 - \frac{k}{m}\right)^n \approx e^{-\frac{kn}{m}} \quad (4)$$

### 3.2. Count-Min

A count-min sketch supports approximate multiset membership queries. It extends the bloom filter bit array structure into a two-dimensional counter array in order to maintain frequency counts.

Let $U$ be the universe of possible set elements with cardinality $n = |U|$ and $M \subset U$ a multiset presented iteratively as a series of updates $(i_t, c_t)$ with $i_t \in M$ and $c_t$ its corresponding count. Although count-min supports counts of any value, we restrict $c_t$ to 1 due to presenting $M$ one element at a time. $M$ can be viewed as a frequency vector $a(T) = [a_1(T), a_2(T), \dots, a_n(T)]$ where, for $t \leq T, i_t = j, a_j(T) = \sum c_t$ denotes the multiplicity of $j$ in $M$ at time $T$. Subsequently we drop $T$ and refer only to current state of $M$.

Based on user defined error margin $\varepsilon$ and failure probability $\delta$, $M$ is represented by a two dimensional array $C$ of width $w = \lceil e/\varepsilon \rceil$ and depth $d = \lceil \ln 1/\delta \rceil$ with all entries initially set to 0. Mapping entries from $U$ to $C$ is achieved via $d$ pairwise independent hash functions $h_1, \dots, h_d: U \to \{1, \dots, w\}$. For each element $x \in M$ and $\forall 1 \leq j \leq d$, $d$ counts from $C$ are incremented by $c_t = 1$: $C[j, h_j(a_x)] = C[j, h_j(a_x)] + 1$.

From a multiset perspective we're only interested in count-min point queries. Given an element $q \in U$, its multiplicity in $M$ is estimated by $\hat{a}_q = \min_{1 \leq j \leq d} C[j, h_j(a_q)]$.

Since we're dealing with non-negative incoming frequency counts ($c_t = 1$) the error due to hash cell collisions is one sided, and, with probability at least $1 - \delta$, $\hat{a}_q \leq a_q + \varepsilon \|a\|_1$.

Again, using the same technique [15] detailed in section 3.1, the requirement for $d$ pairwise independent hash functions can be relaxed. Linear combinations of

$$2\lceil (\ln 1/\delta) / (\ln 1/\varepsilon) \rceil \quad (5)$$

pairwise independent hash functions are used to generate the remaining ones while maintain the same error margin and failure probability.

### 3.3. Compatibility issues

Bloom filter uses mutually independent hash functions while count-min uses pairwise independent ones. As detailed previously in this section, under certain conditions, both requirements can be relaxed to linear combinations of just two independent, uniform random hash functions (Eq. 3). Bloom filter is restricted to disjoint ranges for each hash function. Based on Eq. 5, for $0 < \delta < 1$,

count-min failure probability is lower bounded by its error margin:

$$0 < \varepsilon \leq \delta \quad (6)$$

In what follows we define count-min parameters in terms of optimal bloom filter ones. We start with the bloom filter optimal number of hash functions minimizing *fp* (Eq. 2):

$$k = \frac{ln2}{n} m = \frac{ln2}{n} \cdot -\frac{n \ln(fp)}{ln2^2} = log_2 \frac{1}{fp} \quad (7)$$

Equaling Eq. 7 to count-min depth, for $0 < fp < 1$, count-min failure probability $\delta$ is always lower than *fp*:

$$\delta = fp^{\frac{1}{ln2}} \quad (8)$$

Based on Eq. 7, the range of each bloom hash function becomes:

$$p = \frac{m}{k} = \frac{n}{ln2} \quad (9)$$

Equaling Eq. 9 to count-min width, count-min error margin approaches 0 as *n* increases:

$$\varepsilon = \frac{eln2}{n} \quad (10)$$

Based on Eq. 6, 8 and 10, count-min failure probability is higher than its error margin and no longer lower bounded by it when:

$$fp \geq 2^{\ln\left(\frac{e \, ln2}{n}\right)}, n > e \, ln2 \quad (11)$$

Depending on Eq. 11, $\delta$ is calculated based on bloom filter false positive probability (Eq. 8) or bloom filter number of inserted items (Eq. 10). In the latter case, we only need $n \approx 189$, in order to achieve a count-min failure probability of 1%.

Given a decay rate $\lambda$ for a *d*-dimensional data stream and a minimum density threshold $D_{th}$, the maximum number of generated elements during a cluster dynamic stage becomes (refer to Section 4.2 for further details):

$$n = \frac{1}{2\lambda} \frac{1}{D_{th}} (2d + 1) \quad (12)$$

By constructing the common sketch structure based on optimal bloom filter parameters for a given *n* and *fp*, count-min failure probability and error margin decrease as either the decay rate decreases or the stream dimensionality increases.

## 4. Clustering algorithm

At its core, BloomStream employs a grid based clustering technique. Such techniques discretize the data space into ranges along each dimension mapping the unlimited length data stream into a finite number of grid cells. The number of stream instances within a cell determines its density with neighboring dense cells subsequently aggregated into clusters.

Structures such as hash tables [5, 6], trees [7, 8, 9, 10] allow explicit representation and direct retrieval of dense cells at the cost of dealing with a number of grid cells exponential in the dimensionality of the data stream. By further approximating the data stream with the help of probabilistic data structures a different trade off can be obtained. Space requirements no longer depend on stream dimensionality but on user defined bloom filter parameters as depicted in Section 3.3. Such low space complexity comes at the cost of not being able to directly retrieve dense cells from the corresponding count-min sketch, at least not without being forced to conduct point queries on every possible grid coordinates combination.

This marks a departure from the two phase stream clustering scheme [1] as we lack the capability of easily retrieving the density information in the offline stage. Each data stream instance updates the count-min sketch and, in case the corresponding grid cell is determined to be dense, also updates the existing clustering solution. Cells neighboring the dense grid cell and existing clusters are stored as bloom filters. A series of intersection and reunion operations against these filters creates new clusters, merges or discards existing ones.

### 4.1. Stream density estimate

We assume a data stream of unbounded length composed of *d*-dimensional data instances drawn from $R^d$. We estimate stream density using two summarization techniques. The former discretizes the data while the latter approximates frequency counts over the resulting grid cells. The entire process is conducted under a damped window model.

Definition 1 (Grid Cell). Given a user defined grid resolution *r*, an uniform grid *G* consists of *d* infinite sets of equidistant lines, each parallel to a distinct coordinate axis of $R^d$, where the distance between two neighboring lines is *r*. For a given origin $a = (a_1, a_2, \ldots, a_d)$, *G* represents the set:

$$\{(x_1, x_2, \ldots, x_d) \in R^n : |x_1 - a_1|, \ldots, |x_d - a_d| \in r \cdot \mathbb{Z}\} \quad (13)$$

The above grid partitions $R^n$ into equal-volume, disjoint hyper-rectangular grid cells. Given a data instance $x=(x_1, x_2, \ldots, x_d)$ let $C(x)$ denote its grid cell coordinates represented by the d-tuple $(c_1, c_2, \ldots, c_d)$ where $c_i = \lfloor (x_i - a_i)/r \rfloor$ for $1 \leq i \leq d$.

Definition 2 (Grid Cell Signature). Let $U$ be the spatial domain of all possible grid cell coordinates. All sketches used throughout the proposed algorithm have the same hash table length $m$ and share the same $k$ hash functions $h_1, \ldots, h_k : U \to \{1, \ldots, p\}$ where $p = m/k$ is a prime number (refer to Section 3.3 for further details). Given a data instance $x$, let $CS(x)$ denote its grid cell signature represented by the d-tuple $(cs_1, cs_2, \ldots, cs_k)$ where $cs_i = (i-1) \cdot p + h_i(C(x))$ for $1 \leq i \leq k$.

A single count-min sketch serves as a frequency table of the already discretized data. User defined ε and δ parameters determine the accuracy of point query estimates and the certainty the desired accuracy is reached. Grid cell signatures from each incoming data instance designate $k$ counters from the count-min hash table to be updated. In order to handle concept drift, a timestamp is incorporated into each counter and an exponential decay function is used to assign greater weight to more recent elements during the update process. This is possible because (1) count-min supports fractional weights, and (2) multiplying all count-min counters by a constant $\gamma$ results in accurate point queries scaled by $\gamma$.

Definition 3 (Grid Cell Density). Given a user defined decay rate $\lambda \in (0,1)$, let $t_c$ be the current time and $w(t) = 2^{-\lambda(t_c-t)}$ the weight of a data instance arriving at time $t$. Let *count* denote the count-min hash table and *tcount* the accompanying timestamp data structure. Upon arrival of a new data instance $x$ at time $t$, its corresponding grid cell signature $CS(x) = (cs_1, cs_2, \ldots, cs_k)$ designates $k$ count-min counters to be updated, $1 \leq i \leq k$:

$$count[i, cs_i] = w(tcount[i, cs_i]) \cdot count[i, cs_i] + 1 \quad (14)$$

Each time a counter is updated, its corresponding timestamp is also updated to the current time. Following the sketch update, the grid cell density can now be computed as $D(C(x), t) = \min_i count[i, cs_i]$. A grid cell is said to be dense if its estimated density is above a user defined threshold $D_{th}$.

### 4.2. Cluster membership estimate

A dense cell is similar to a core point from traditional density based clustering algorithms. All points in the neighborhood of a core point $q$ are said to be directly density-reachable from $q$ and belonging to the same cluster [16]. A dense cell and its neighboring cells always belong to the same cluster. A clustering update is triggered each time such a cell is found following the arrival of a new data instance. It does not matter if the neighboring cells are dense or not. If they are identified later on as dense, they will trigger additional clustering updates.

Definition 4 (Grid Cell Neighborhood). Grid cells $C_i$ and $C_j$, with coordinates $(i_1, i_2, \ldots, i_d)$ and $(j_1, j_2, \ldots, j_d)$ respectively, are neighbors if:

$$\begin{cases} \exists l, 1 \leq l \ll d, \ |i_l - j_l| = 1, and, \\ \forall h \neq l, 1 \leq h \leq d, \ i_h = j_h \end{cases} \quad (15)$$

Thus each cell has $2d$ orthogonal neighbors. Given a grid cell $C_x$, let $CN(C_x)$ stand for its corresponding grid cell neighborhood consisting of $2d + 1$ cells, the cell itself and its neighbors.

The proposed algorithm uses bloom filters to represent individual clusters. Each bloom filter encodes the corresponding cluster grid coordinates sharing the same structure size and hash functions with the count-min sketch detailed in the previous section. This allows the grid cell signature used for updating count-min to also be used for updating bloom filters.

Definition 5 (Grid Cluster Fragment). Given a grid cell $C_x$, for each neighboring cell $C_i \in CN(C_x)$ with signature $CS_i = (cs_{i1}, cs_{i2}, \ldots, cs_{ik})$, $1 \leq i \leq 2d + 1$, we construct a partitioned bloom filter $BF_i$ using a bit vector of $m$ bits and $k$ hash functions. Each filter encodes coordinates for a single cell, $BF_i[cs_{ij}] = 1$ for $1 \leq j \leq k$. Let $GCF(C_x) = \{BF_i : 1 \leq i \leq 2d + 1\}$ stand for the grid cluster fragment of grid cell $C_x$.

Definition 6 (Grid Cluster Signature). Let $GCF_1, \ldots, GCF_n$ represent $n$ grid cluster fragments with at least one element in common, $\cap_{i=1}^n GCF_i \neq \emptyset$. Let $BF_{ij}$ representing the $j^{th}$ partitioned bloom filter of the $i^{th}$ cluster fragment, $1 \leq j \leq 2d + 1$. A grid cluster signature $GCS$ is represented by a partitioned bloom filter generated from the union of all bloom filters from each cluster fragment, $0 \leq b < m$:

$$GCS[b] = OR_{i=1}^n BF_{ij}[b] \quad (16)$$

Storing clusters as bloom filters presents unique challenges under the data stream model. Although simple to construct and fast to update, the partitioned bloom filter is designed for static datasets, supporting only insert and query operations. It does not support delete operations nor does it contain information supporting a cluster split. Various intra-cluster summarized statistics used for defining a micro-cluster maximum boundary [1], a cluster split confidence level given by the Hoeffding bound [17], etc., are just not available when employing bloom filters.

Definition 7 (Grid Cluster). Given a grid cluster signature $GCS_x$, let $t_x$ represent its creation time and $t_c$ the current time. Let $T_{th} = 1/\lambda$ denote a time threshold equal to the half-life of the exponential decay function used for assigning weights to count-min counts. A grid cluster is defined as a 2-vector $GC = (GCS_x, t_x)$ and said to be dynamic if $t_c - t_x < T_{th}/2$ or stable if $T_{th}/2 \leq t_c - t_x < T_{th}$. Clusters where $t_c - t_x > T_{th}$ are labeled as expired and removed.

The periodic cluster removal conditioned by $T_{th}$, ensures each bloom cluster signature represents grid cells with, at minimum, half the density when they were first added to cluster. This keeps cluster representation up-to-date effectively negating the need for splitting in case of concept drift.

Assuming constant high density for a given grid cell, as one cluster is removed, a new one emerges sharing the same grid cell. It is important data instances from both clusters share the same cluster label as the data stream exhibits only one real cluster. A time overlap allows for such a connection linking dynamic clusters to stable ones. The proposed approach resembles bloom active-active buffering [4] where two bloom filters are linked together. In our case both are used for membership testing but only one is used for receiving new data instances. A dynamic cluster is trend-setting, allowing expansion and mergers. A stable cluster no longer supports updates but takes part in the process of deciding which cluster label a given data instance belongs to.

| Algorithm 1: getMatchingClusters(clustFragment) |
|---|
| 1:    matchingClusters = new empty list |
| 2:    **foreach** bloomFilter **in** clustFragment |
| 3:       **foreach** cluster **in** clustering solution |
| 4:          clustSignature = signature for cluster |
| 5:          **if** bloomFilter matches clustSignature **then** |
| 6:             add cluster to matchingClusters |
| 7:    **return** matchingClusters |

Prior to a clustering update triggered by a dense cell, we filter existing clusters based on the cell's cluster fragment. Each bloom filter from the fragment is tested for membership in all available cluster signatures. Clusters with matching signatures are then subject to the clustering update procedure. Algorithm 1 shows the pseudo-code for the signature matching algorithm.

| Algorithm 2: clustUpdate(clustFragment, matchingClusters) |
|---|
| 1:    newClust = create new cluster |
| 2:    newSignature = signature for newClust |
| 3:    **foreach** bloomFilter **in** clustFragment |
| 4:       add bloomFilter to newSignature |
| 5:    dynamicLabels = empty list |
| 6:    stableLabels = empty list |
| 7:    **foreach** oldClust **in** matchingClusters |
| 8:       oldSignature = signature for oldClust |
| 9:       **if** cluster is dynamic **then** |
| 10:          add oldSignature to newSignature |
| 11:          add oldClust label to dynamicLabels |
| 12:          remove oldClust |
| 13:       **else if** cluster is stable **then** |
| 14:          link oldClust to newClust |
| 15:          add oldClust label to stableLabels |
| 16:       **else** |
| 17:          remove oldClust |
| 18:    commonLabels = intersect dynamicLabels, stableLabels |
| 19:    **if** commonClustLabels not empty **then** |
| 20:       newLabel = random entry from commonLabels |
| 21:    **else if** stableLabels not empty **then** |
| 22:       newLabel = random entry from stableLabels |
| 23:    **else if** dynamicLabels not empty **then** |
| 24:       newLabel = random entry from dynamicLabels |
| 25:    **else** |
| 26:       newLabel = generate new cluster label |
| 27:    assign newLabel to newClust |
| 28:    add newClust to the clustering solution |

During each signature matching we conduct $2d + 1$ queries against existing grid cluster signatures. These bloom filter operations are the computationally most expensive part of the algorithm. In order to accelerate queries we store the grid cluster signatures in a data structure similar to a flat bloom filter index [18]. Our version of the index is implemented as a single array of size *m*, where each element is represented by a bit array equal in length with the total number of current clusters. Thus, the $j^{th}$ bit from the $i^{th}$ array position represents the $i^{th}$ bit of the $j^{th}$ cluster signature. Given that a bloom filter from a cluster fragment has *k* bits set, we can test if it matches any of the existing cluster signatures in at most *k* bitwise AND operations.

A clustering update takes as input a dense cell's cluster fragment and its matching clusters. Initially, a new cluster is created and all bloom filters from the cluster fragment are added to it. The newly created cluster represents an entirely new cluster if no matching clusters are present, a cluster expansion if a single dynamic cluster is present or a cluster merger if multiple dynamic clusters are present. Both cluster expansion and merger are conducted by computing the bloom filter union between the corresponding cluster signatures. If one or multiple stable clusters are present, the new cluster links to them without performing a bloom filter union. All linked clusters return the same label. The label for the newly created one is inherited from one of the matching clusters with stable clusters taking precedence over dynamic ones in order to ensure label continuity. Only when no matching clusters are present is the new cluster assigned a new label. It is at this point that we also perform cluster removal for all matching but expired clusters. The entire process is illustrated in Algorithm 2 pseudo-code.

## 5. Experimental results

A thorough experimental evaluation comparing BloomStream scalability and accuracy against D-Stream II is presented in this section. BloomStream is implemented in C++ with an R interface for the Stream R package [19]. We use the D-Stream II C++ implementation from the same package. This allows us to test and evaluate both algorithms under the common Stream framework. All experiments are conducted on a 3.2GHz Intel Core i5 with 8 GB memory running Ubuntu 16.04 LTS.

### 5.1. Evaluation and data sets

BloomStream does not differentiate between typical online and offline clustering phases and does not employ any distance based summaries. This rules out all evaluation measures at microcluster level and all distance based evaluation measures at macro-cluster level. From the remaining external validation measures at macro-cluster level, we've selected purity to quantify clustering accuracy against ground truth.

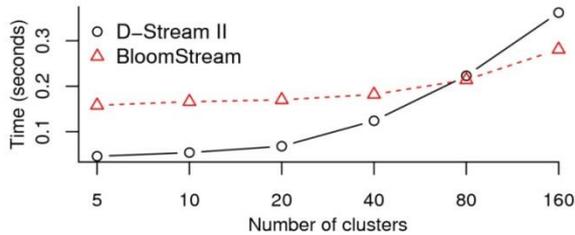

**Fig. 1.** Execution time vs. number of clusters.

Let $C = \{c_1, c_2, ..., c_k\}$ represent a set of clusters and $L = \{l_1, l_2, ..., l_j\}$ a set of ground truth labels where $c_k$ and $l_j$ denote the set of data instances in $c_k$ and $l_j$ respectively. Taking values in the unit range [0, 1], purity represents the average purity of clusters with respect to the ground truth:

$$Purity = \frac{1}{k} \sum_{i=1}^{k} \frac{max_j|c_i \cap l_j|}{|c_i|} \quad (17)$$

Scalability tests are conducted against synthetic data streams containing a series of Gaussian clusters with 10% uniformly distributed noise. The clusters are well separated with a minimum distance between any two cluster centers of at least 4. We start with a 5 dimensional data stream containing 5 clusters across a 2k window length and separately vary the number of dimensions and clusters from 5 to 160. We set the evaluation horizon equal to the window length.

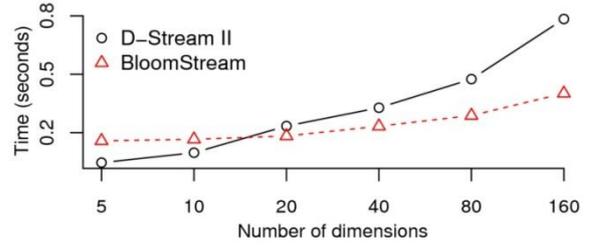

**Fig. 2.** Execution time vs. stream dimensionality.

Accuracy tests are conducted against KDD CUP'99 Network Intrusion Detection dataset, consisting of approximately 5 million TCP connection records, with each record containing 42 attributes. There are 25 clusters present representing 24 attack types in addition to normal connections. Due to the sporadic nature of the attacks, the corresponding data stream is rapidly evolving with a varying number of clusters at any given time, thus we set the evaluation horizon to 1000. As in [5, 7], we use all 34 continuous attributes, normalized to [0, 1].

### 5.2. Scalability and accuracy results

Throughout all scalability experiments, both algorithms achieve a purity value of at least 0.9 using the same fixed configuration. With the exception of grid size, all D-Stream II parameters match the adopted values from the corresponding paper. We set the same grid size of 1.5 to both algorithms. BloomStream density threshold $D_{th}$ is set to 3 and its decay rate $\lambda$ to 0.001, the D-Stream II corresponding value. Finally, BloomStream hash table length $m$ and number of hash functions $k$ are set to 70063 bits (8.56 KB) and 7 respectively, corresponding to a bloom filter capacity of 6935 elements while maintaining a false positive rate of ~0.78%.

The above bloom filter parameters have been set based on Eq. 12. For $\lambda = 0.001$, $D_{th} = 3$ we can encounter up to 166 dense cells during a cluster dynamic stage. A cluster always contains the dense cell neighborhood of $2d + 1$ cells, not just the dense cell itself. Assuming all cells have a density equal to $D_{th}$ and belong to the same cluster, the bloom filter representing the cluster requires a capacity $n = 1826$ for $d = 5$, $n = 53286$ for $d = 160$. In practice, the above scenario is highly unlikely and we set the bloom filter capacity to fraction of $n$ between 10-15%.

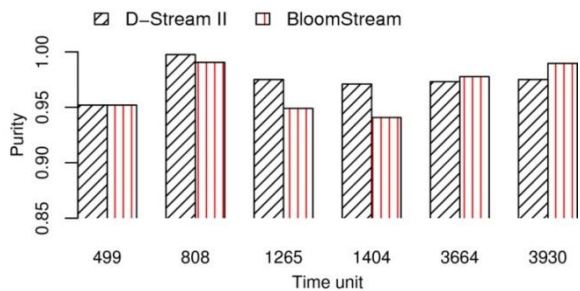

**Fig. 3.** Clustering quality on Network Intrusion dataset

With the help of the count-min sketch, BloomStream determines if a new data instance belongs to a dense grid cell in $O(k)$ time. If the cell is found to be dense, a bloom filter from the corresponding cluster fragment completes a clustering update in $O(m)$ time. Total clustering model update time depends on the number of dense cells and grid cluster fragment size (Eq. 12). Evaluating a data instance in order to determine its cluster label takes $O(m)$ time, the time it takes to test if the corresponding cell signature matches any of the existing cluster signatures; it depends neither on number of clusters nor stream dimensionality.

BloomStream scalability regarding number of dense cells and grid cluster fragment size is measured by varying number of clusters (Fig. 1) and number of dimensions (Fig. 2) respectively. Note the logarithmic scale on X axis. The recorded execution time includes both the clustering update and clustering evaluation time. In both experiments, BloomStream evaluation time is constant and represents the majority of the total execution time when either the number of clusters or stream dimensionality is low. This accounts for the approximately constant execution time when the synthetic data stream has a dimensionality lower than 20 or contains less than 40 clusters. In these scenarios, D-Stream II execution times are lower as its clustering and evaluation time scale with the data with no initial overhead.

D-Stream II scalability is affected by the fact it needs to specifically handle sporadic or transitional grid cells due to noise or emerging clusters. In addition, its grid cell density threshold is based on average grid cell density approaching 0 as the number of dimensions increases, thus resulting in a large number of micro-clusters. In contrast, BloomStream does not have any of these problems, scaling better as the number of clusters or stream dimensionality increases.

We evaluate BloomStream clustering quality in comparison with D-Stream II using the Network Intrusion dataset. With an evaluation horizon set to 1000, we only consider time units containing less than 50% normal connections with a maximum or near maximum number of different clusters. Following this approach, the selected time units illustrated in Fig. 3 contain 3 or 4 different clusters. All D-Stream II parameters match the adopted values from the corresponding paper. BloomStream parameters remain the same with the exception of grid size now set to 0.03. Both clustering algorithms achieve an average purity of over 0.95 with each algorithm achieving higher purity depending on the selected time unit.

## 6. Conclusions

By discretizing the data stream into grid cells, grid based stream clustering algorithms are required to handle an exponentially increasing number of grid cells as the stream dimensionality grows. Existing algorithms try to mitigate this problem with various degrees of success by implementing special structures and logic in order to handle sporadic or transitional grid cells due to noise, outliers or emerging clusters. The proposed algorithm, BloomStream, circumvents this problem by estimating grid cell density and not explicitly storing it.

Given a data stream instance, BloomStream clustering update procedure is only triggered when the corresponding grid cell is found to be dense following a count-min update taking constant time, independent of stream dimensionality. The same set of hash functions used for updating the count-min sketch is also used for updating a series of bloom filters representing the identified clusters. Clusters are created, merged or discarded via intersection and reunion operations against these filters. The use of the above probabilistic data structures allows for a user defined trade-off, in terms of probabilistic accuracy guarantees, between stream clustering speed and accuracy.